%% file: refine2013_zeyda.tex
\documentclass{eptcs}
\usepackage{wasysym}
\usepackage[pdftex]{graphicx}
\usepackage[colour]{circus}
\usepackage{zeyda}

\hypersetup{colorlinks=true,linkcolor=OliveGreen,citecolor=Red,urlcolor=Blue}

\newcommand{\keywordname}{\textbf{Keywords:}}

\newcommand{\figname}{Fig.\thinspace}
\newcommand{\lawname}{Law~}

\title{Refining SCJ Mission Specifications into\\Parallel Handler Designs}

\author{Frank Zeyda
\institute{University of York, York, YO10 5GH, UK.}
\email{frank.zeyda@york.ac.uk}
\and Ana Cavalcanti
\institute{University of York, York, YO10 5GH, UK.}
\email{ana.cavalcanti@york.ac.uk}
}

\def\titlerunning{Refining SCJ Mission Specifications into Parallel Handler Designs}
\def\authorrunning{F.~Zeyda \& \hspace{-1.5ex} A.~Cavalcanti}

\begin{document}

\maketitle


\begin{abstract}
Safety-Critical Java~(SCJ) is a recent technology that restricts the execution and memory model of Java in such a way that applications can be statically analysed and certified for their real-time properties and safe use of memory. Our interest is in the development of comprehensive and sound techniques for the formal specification, refinement, design, and implementation of SCJ programs, using a correct-by-construction approach. As part of this work, we present here an account of laws and patterns that are of general use for the refinement of SCJ mission specifications into designs of parallel handlers used in the SCJ programming paradigm. Our notation is a combination of languages from the {\Circus} family, supporting state-rich reactive models with the addition of class objects and real-time properties. Our work is a first step to elicit laws of programming for SCJ and fits into a refinement strategy that we have developed previously to derive SCJ programs.
\end{abstract}

\keywordname~SCJ, models, refinement, laws, patterns, tactics, {\Circus}.


\section{Introduction}
\label{sec:Introduction}

Java is indisputably one of the most popular programming languages. Despite this, its use in the safety-critical industry has been modest due to Java's generality and rich set of features. Significant issues are, for example, the use of garbage collection and problems related to thread prioritisation~\cite{STR06,JSR302}, which render it inadequate for time-critical applications. Safety-Critical Java~(SCJ)~\cite{HHLNSV09}, a recent initiative, addresses these issues by introducing a restricted subset of Java; it is based on the Real-time Specification for Java (RTSJ)~\cite{Wel04}, but further restricts RTSJ's execution and memory model. This facilitates the formal analysis of SCJ applications, and thereby enables the application of formal methods to satisfy stringent criteria of certification standards like DO-178C.

SCJ is organised in three levels (Level 0 to Level 2) that define progressively more complex models of execution. Our focus is SCJ Level 1, which roughly corresponds to the Ravenscar profile for Ada~\cite{Bur99}. At Level 1, applications are organised as a sequence of missions, and each mission consists of a set of handlers that are executed in parallel. Handlers can either be periodic, which means they are released at regular time intervals, or aperiodic implying that they are released sporadically by some external event or stimulus. When a handler is released, its \verb"handleAsyncEvent()" method is scheduled for execution.

Our previous work has focused on complementing the informal account of SCJ~\cite{JSR302} with a formal model of SCJ's mission-based execution paradigm~\cite{ZCW11} and memory model~\cite{CWW11}. Our notation is a combination of languages from the {\Circus} family~\cite{CSW03,CSW05,SCJS09}, specifically tailored for the specification and development of state-rich reactive systems with the addition of discrete time, object-orientation, and object references. We have also proposed a refinement strategy~\cite{CZWWW12} to transform abstract specifications of SCJ programs into models that directly correspond to SCJ programs. Such a strategy is inherently ambitious and complex, as it simultaneously addresses a multitude of concerns. Therefore, it is not surprising that the existing work~\cite{CZWWW12} only gives a broad description of the top-level approach; details of the application of this strategy to a specific example are available in~\cite{ZCWWW12}.

Our contribution in this paper is to examine in detail the refinement of centralised and sequential specifications of missions into parallel handler designs. Our general starting point is a {\Circus} process specification that supports all constructs of {\Circus}, including Z data operations, classes, and Timed CSP constructs, except for parallelism and interleaving. We then show how decomposition at the level of data operations, time budgets, and process actions can be used to transform the model into a uniform shape that determines the structure and behaviour of handlers of an SCJ mission. Refinement laws directly reflect particular program designs that encapsulate the way in which data is shared and how the computational work is divided between the handlers of a mission.

The motivation for our work is to pave the way for automated tool support. Due to the novelty of SCJ, there are not many tools currently available that support the development of critical software in SCJ. The available tools mostly focus on isolated statically-checkable properties~\cite{TPV10,DHS12,HL11}, but do not address the combination of concerns that characterise the SCJ paradigm. While we do address many concerns of SCJ simultaneously by using a highly expressive language, the practicalities of performing actual refinements are largely an open problem. It is, clearly, unrealistic to carry out such refinements entirely by hand, which is well illustrated by the complexity of the example in~\cite{ZCWWW12}. Some refinement steps are, however, inherently difficult to automate. Our work, most importantly, highlights where automation is feasible, and where human guidance is indispensable to guide the refinement process.

The results in this paper contribute towards elaborating the proposed refinement strategy for SCJ in~\cite{CZWWW12}, but they are also useful outside the context of that technique. Decomposition of centralised models is a general issue in refinement-based techniques~\cite{CSW03}, and the models we produce can, in principle, serve as a starting point for any form of parallel implementation. As the essence of the SCJ paradigm~(its mission-based execution model) can be captured independently of the Java language, our account on mission decomposition is relevant for other languages that adopt a similar execution model, too.

The structure of this paper is as follows. In Section~\ref{sec:Preliminaries} we review preliminary material:~Safety-Critical Java and the {\Circus} family of languages. Section~\ref{sec:Strategy} then discusses our refinement laws, and Section~\ref{sec:Example} presents an example of their application. Finally, in Section~\ref{sec:Conclusion} we conclude and suggest future work.


\section{Preliminaries}
\label{sec:Preliminaries}


We here discuss in more detail Level 1 SCJ and the {\Circus} family of notations.

\subsection{Level 1 SCJ}
\label{sec:SCJ}

The execution model for SCJ Level 1 programs is based on four primary conceptual entities:~safelet, mission sequencer, missions and handlers. They are realised by classes that derive either from an interface or abstract class of the SCJ API. Namely, these are \code{Safelet}, \code{MissionSequencer}, \code{Mission}, \code{Periodic\-Event\-Handler}, and \code{AperiodicEventHandler}.

\begin{figure}[t]
\center
\includegraphics*[scale=0.75,trim=20 335 20 345]{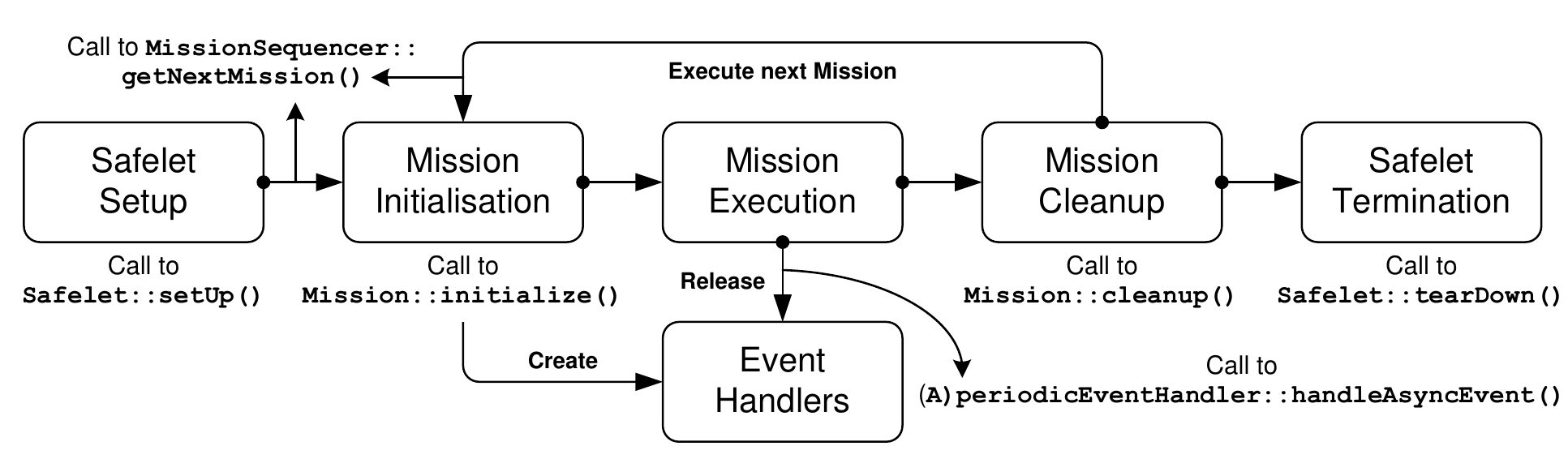}
\vspace{-8pt}
\caption{Life-cycle of a safelet during execution of a Level 1 application}
\label{fig:SafeletLifecycle}
\end{figure}

\figname\ref{fig:SafeletLifecycle} illustrates the life-cycle of a Level 1 safelet, the top-level entity of an SCJ application. The SCJ infrastructure\footnote{By `SCJ infrastructure' we mean an SCJ-compliant virtual machine.} first initialises the safelet. This is followed by a series of mission executions, each involving the initialisation, execution and termination of a particular mission of the safelet. Mission initialisation creates the mission's event handlers, which are released either periodically or by external events during mission execution. When there are no more missions to execute, the safelet terminates.

In terms of the SCJ API, a class implementing \code{Safelet} has to provide the methods \code{setUp()} and \code{tearDown()}, which are called by the SCJ infrastructure to initialise and shutdown the safelet. Another method~(not in \figname\ref{fig:SafeletLifecycle}) is called on the safelet object to obtain the mission sequencer of the application, which defines the sequence of missions to execute. In addition, various methods are called by the infrastructure on the mission sequencer, mission and handler objects during execution of the safelet. Most notably, these are \code{getNextMission()} to obtain the next mission to execute, \code{initialize()} to create the handlers of a mission, and \code{handleAsyncEvent()} when a handler is released. An SCJ program must provide implementations of these methods, and it thereby defines the architecture of the application in terms of missions and handlers. (We note that although the missions and handlers of a safelet are determined at run-time, we assume in our model that they are \textit{a priori} fixed.)

When a mission terminates, \code{cleanup()} is called on the mission object to perform application-specific clean-up tasks. As already mentioned, the entire safelet terminates when there are no more missions to execute, signalled by \code{getNextMission()} returning a \code{null} reference. In summary, the safelet and the mission sequencer are control components that orchestrate the execution of the missions (and their handlers). The missions and the handlers, on the other hand, are the central components that implement the behaviour of the program, and the main focus of our work here.

\subsection[The Circus family]{The {\Circus} family}
\label{sec:Circus}

{\Circus}~\cite{CSW03} is a language for specification and refinement of state-rich reactive systems. It combines notations from CSP~\cite{Ros97}, Z~\cite{WD96}, and Morgan's refinement calculus~\cite{Mor90}. As in CSP, the key elements of {\Circus} models are processes that can interact with their environment through channels. Unlike CSP, {\Circus} processes encapsulate a state that can be modified by actions and data operations of the process. {\Circus} has a denotational semantics defined using the Unifying Theories of Programming~\cite{OCW09}.

\begin{figure}[t]
\centering
\fbox{
\begin{minipage}{0.9\textwidth}
\vspace{-1em}
\input{Target.tex}
\vspace{-1.5em}
\end{minipage}}
\caption{Target for refinement transforming mission models.}
\label{fig:Target}
\end{figure}
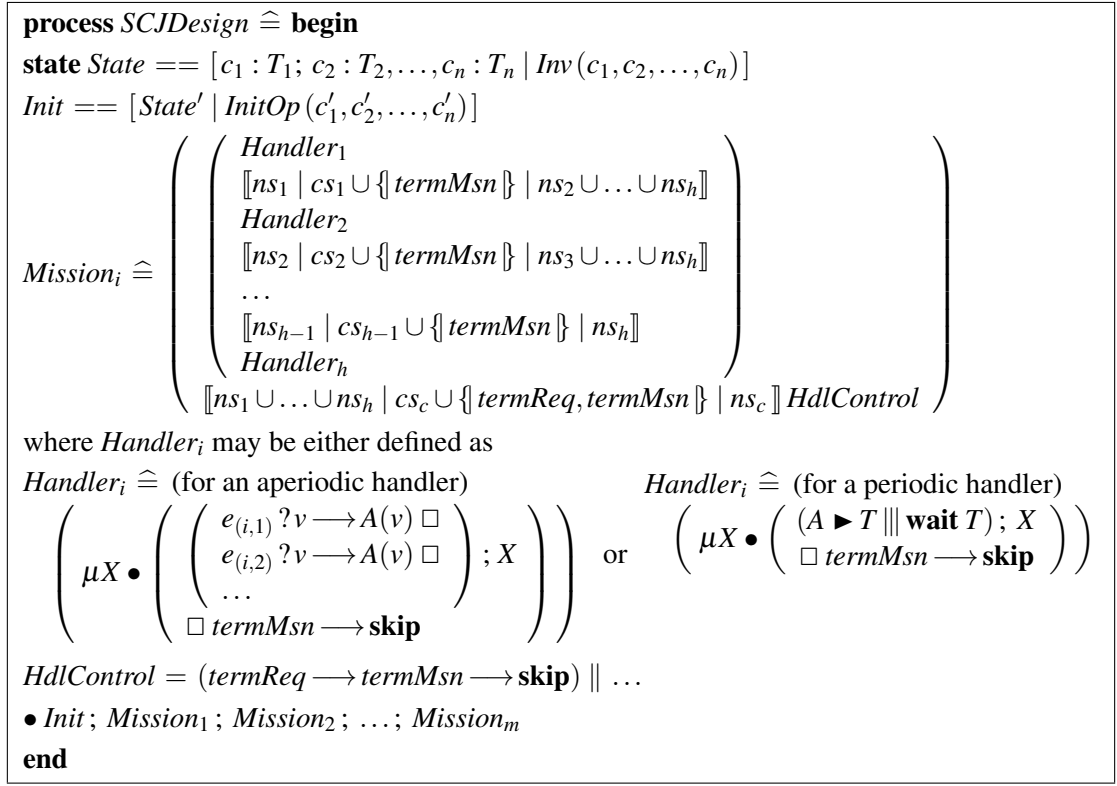

An example of a {\Circus} process is given in \figname\ref{fig:Target}. It illustrates the general form of an SCJ handler design, and the laws we discuss in the next section transform (sequential) specifications of safelets into processes of this shape. The name of the process is $SCJDesign$, and its state is defined by the $State$ schema, introducing the components $c_i$ of type $T_i$~($Inv$ is an optional state invariant). The $T_i$ may be Z schema types or {\OhCircus} class types, as it is also the case for the types in any of the laws. We then have local action definitions for $Init$, $Mission_i$, $Handler_i$ and $HdlControl$. The actual behaviour of the process is defined by the main action at the bottom after the `$\circspot$'; it typically makes use of the local actions.

Local actions can be either specified by Z operation schemas or using a mixture of CSP constructs and guarded commands. Here, $Init$ is a Z operation that initialises the state, and $Mission_i$ and $Handler_i$ are CSP actions that provide models of missions and handlers as they emerge during verification. Each $Mission_i$ action is defined by a parallel composition of a mission-specific set of handler actions. In {\Circus}, parallel composition of two actions $A_1$ and $A_2$ is written as $A_1 \lpar ns_1 | cs | ns_2 \rpar A_2$, where $cs$ is a set of interface channels that require synchronisation of the actions, and $ns_1$ and $ns_2$ are disjoint sets of variables that each action is allowed to write to. Hence, all handlers of a mission write to mutually disjoint parts of the state space, determined by the $ns_i$. This ensures that all {\Circus} constructs (including parallel composition) are monotonic with respect to refinement due to intrinsic non-interference in shared data access by parallel processes and actions.

The action $HdlControl$ is included to incorporate control mechanisms. It controls termination of the mission via the channels $termReq$~(for a termination request raised by one of the handlers) and $termMsn$~(to synchronously terminate the handlers). It also permits the definition of additional control actions~(dots) whose design is not a concern for mission decomposition into handlers.

The handler models, captured by the actions $Handler_i$, take different shapes for aperiodic and periodic handlers. Both, however, have the form of a recursion $(\circmu X \circspot A \circseq X \extchoice ~ termMsn \then \Skip)$ that repetitively executes some action $A$ and at the same time enables termination by $HdlControl$. Aperiodic handlers are modelled by an external choice that synchronises on a set of channels $e_{(i, 1)}$, $e_{(i, 2)}$, and so on, which correspond to SCJ events that are bound to the handler $i$ and therefore cause its release. Potentially, each event provides an input $v$, and the \code{handleAsyncEvent()} method is specified by $A(v)$.

For periodic handlers, the repetitive behaviour is determined by the action $A \circdeadlineterm T ~ \interleave ~ \circwait ~ T$, using additionally constructs from {\CircusTime}. The $A \circdeadlineterm T$ operator imposes a termination deadline $T$ on $A$, $\circwait ~ T$ corresponds to a delay of $T$ time units, and the interleaving with $\circwait ~ T$ prevents the action from terminating \emph{before} $T$ time units have elapsed. Hence, we obtain a cyclic behaviour that executes the \code{handleAsyncEvent()} method $A$ once every $T$ time units. For clarification, we point out that all {\CircusTime} constructs take relative times as their arguments.

We note that interleaving~($A_1 \interleave A_2$) is a special case of parallelism where the synchronisation set $cs$ is empty; termination only occurs when both parallel actions have terminated. Later on, we also make use of the $\circwait ~ t_1 \upto t_2$ statement, which corresponds to a nondeterministic delay between $t_1$ and $t_2$ time units, and $A \circdeadlinesync ~ T$ which is a deadline on $A$ to interact via a visible event, such as a communication or synchronisation.

The main action of $SCJDesign$ at the end first initialises the state and then executes all missions in sequence~(operator~$A_1 \circseq A_2$). In \figname\ref{fig:Target}, we use notations from both {\Circus} and {\CircusTime}~\cite{SCJS09}, and generally also support the use of constructs from {\OhCircus}~\cite{CSW05} for class objects. The UTP~\cite{HH98} enables us to give a sound semantic foundation to this combination of languages.


\section{Refinement Laws}
\label{sec:Strategy}

Our starting point is a centralised mission specification that defines communication patterns, data operations, and timing restrictions using sequential {\Circus} actions. We deal with three aspects of the verification of a mission implementation with respect to such a centralised specification. The first aspect is decomposition of data operations to introduce functional models of handlers. The second is distribution of time budgets between the handlers. And the third is parallelisation of handlers to match the architecture of Level 1 SCJ; this also addresses data flow and control mechanisms via communications. We present here collections of {\Circus} refinement laws and tactics to address each of these verification issues. Although some of these laws have already been given in~\cite{ZRC} and \cite{CSW03}, the parallelisation laws in \figname\ref{fig:ParPattn2} and \figname\ref{fig:conj-to-par-2} are to our knowledge novel, and so are the {\CircusTime} laws in Section~\ref{sec:Stage2}.


\subsection{Decomposition of data operations}
\label{sec:Stage1}

Here we target data operations. We note that we do not generally require that the specification of a mission involves a single data operation. For missions with simple interaction patterns, such as reading an input, performing a computation, and writing an output, it is possible to capture the functional aspects of the mission in a single data operation. In the general case, however, where inputs and outputs may occur sporadically during mission execution, a functional mission model may be split into more than one data operation. We assume, on the other hand, that all data operations specify mission behaviour at a suitably high level of abstraction:~this means they are centralised models of functionality, and hence do not already encapsulate any form of computational design.

Our goal is to decompose data operations so that the (functional) specifications of individual handlers emerge. We employ schema composition to model sequential execution of handlers, and schema conjunction to model parallel execution of handlers. All refinement is carried out at the level of Z. The Z Refinement Calculus~\cite{ZRC,Gro02}, whose laws are valid in {\Circus}~\cite{OCW09}, provides the foundation for our laws here. The laws we present are therefore applicable and relevant for Z refinement in general.

Though~\cite{ZRC,Gro02}, for example, present a collection of laws that address issues of decomposition too, it is well understood that decomposition of data operations is overall difficult to\commentout{systematise and} automate. We propose a number of specialised laws that cover a broad spectrum of mission designs. Each law encapsulates either a sequential or parallel design that carries out a centralised computation by two or more handlers.

\paragraph{Laws for sequential decomposition of data operations}

We distinguish two fundamental cases. The first one assumes no dependency between the data operations in terms of the computed results. The corresponding law is presented in~\figname\ref{fig:SeqPattn1}. We assume the existence of a $State$ schema that specifies the state on which the operations act. It is partitioned into two disjoint lists of variables, $x$ and $y$, which are respectively constrained by the state invariants $I_1(x)$ and $I_2(y)$. The law decomposes $Op$ into a sequence $Op_1 \semi Op_2$, where $Op_1$ only modifies the components in $x$, and $Op_2$ only modifies the components in $y$ and does not depend on $x$. Application of this law entails transforming the predicate of an operation schema into a form $P(x, y, x') \land Q(y, y')$. This may in general require intelligent decision making, but in some cases ought to be automatable using elementary laws and syntax-driven rewriting realised by generic tactics of proof for the Z mathematical notation.

\begin{figure}[t]
\begin{law}
\input{SeqPattn1.tex}
\label{law:SeqPattn1}
\end{law}
\vspace{-0.5em}
\caption{Sequential decomposition of independent data operations.}
\label{fig:SeqPattn1}
\end{figure}

The second case is where there exists a dependency between the data operations in terms of the result. That is, the second operation uses data that is computed by the first one. Here, we have the general law in \figname\ref{fig:SeqPattn2}. The crucial difference is in the shape of the predicate of the refined operation $Op$, where $Q(x', y, y')$ refers to the final value of $x$. The state invariant is decomposed as well, namely into a conjunct $I_1(x)$ that only considers constraints on $x$, and another conjunct $I_2(x, y)$ that relates $x$ and $y$. The invariant $I_2(x, y)$ is not enforced by the $\Delta$ and $\Xi$ schemas but moved into the predicates for technical reasons: we observe that $I_2$ may in fact not hold for the intermediate state of the decomposition.

\begin{figure}[t]
\vspace{2em}
\begin{law}
\input{SeqPattn2.tex}
\label{law:SeqPattn2}
\end{law}
\vspace{-0.5em}
\caption{Sequential decomposition of dependent data operations.}
\label{fig:SeqPattn2}
\end{figure}

The propagation of invariants proves to be especially important to facilitate further decomposition and later algorithmic refinement. Invariant decomposition once again requires guidance. It involves the transformation of a single invariant $I(x, y)$ into the conjunction $I_1(x) \land I_2(x, y)$ so that all relevant knowledge about the components in $x$ is encoded by $I_1(x)$.

We have defined several variations of the previous two laws that moreover deal with inputs and outputs of operations. We omit their discussion as they are straightforward generalisations. They can, however, be found in the appendix of~\cite{CZWWW12}. Next, we take a look at parallel decomposition.

\paragraph{Laws for parallel decomposition of data operations}

As before, we have a pair of laws that consider the case of independent and dependent data operations. Dependency here means that the operations cumulatively participate in the computation of some result. For independent data operations, the law is similar to that in \figname\ref{fig:SeqPattn1} with a small modification of the right-hand side: firstly, the sequence $Op_1 \semi Op_2$ is replaced by a conjunction $Op_1 \land Op_2$, and secondly, we remove the $\Xi$ schemas in the declaration part of $Op_1$ and $Op_2$. The fact that both laws have the same left-hand side illustrates that there is often more than one possible handler design, giving rise to different degrees of parallelisation.

A more interesting parallelisation law is presented in \figname\ref{fig:ParPattn2}. There, we have $n$ handlers participating in the computation of the result $r$ and using the components $x$. The behaviour of the handlers is specified by the predicate $Q(r_i, i, x)$ for $1 \le i \le n$. Decomposition here yields a conjunction that includes a conjunct $POp$ for each handler, as well as a merge operation $MOp$ that collects the partial results $r_i$ to compute the overall result of the refined operation. Following the Z convention, the symbols `$?$' and `$!$' in the declaration part of the schemas $POp$ and $MOp$ are used to identify input and output parameters. The merge operation is parametrised by a bag to enforce syntactically that the order in which the results are delivered is irrelevant. Hence, we require that the binary operation used in the merge is associative and commutative; the merge basically consists of folding this operation over the list of partial results.

\begin{figure}[t]
\begin{law}
\input{ParPattn2.tex}
\label{law:ParPattn2}
\end{law}
\vspace{-1em}
\caption{Parallel decomposition of dependent data operations.}
\label{fig:ParPattn2}
\end{figure}

It turns out that the application of the above decomposition laws, in comparison to subsequent sets of laws, is the most challenging to automate. The developer needs to determine the target of each law application, that is, the schema predicates on the right-hand side of the laws. With that, a verification condition can be generated to establish that the predicate of the schema being refined can be written in the form required for the application of the law. Specialised proof tactics will be useful in this context.


\subsection{Distribution of time budgets}
\label{sec:Stage2}

Data operations in {\Circus} are atomic and instantaneous. Hence, all timing behaviour has to be specified explicitly using timed action operators. Time budgets specify the permissible amount of time that an implementation may take to execute a data operation; in {\Circus}, they can be captured by nondeterministic wait statements of the form $\circwait 0 \upto t$ that precede a data operation. The laws in this section are hence essentially about $\circwait$ statements modelling time budgets, and, therefore, are useful in any context where we want to reason about the timing of Z data operations in (\textsf{\slshape Oh}){\Circus}.

Our general assumption is that the specification of mission behaviour may utilise $\circwait$ statements in arbitrary places. The laws in this section decompose and distribute those $\circwait$s in order to attach them to the data operations emerging from decomposition in the previous section. Using these laws, we can equip each decomposed data operation $Op$ with an operation-specific time budget $\circwait 0 \upto TB_{Op}$, where $TB_{Op}$ determines the amount of time the operation may take to execute in the SCJ program.

The refinement laws needed can be divided into two classes. In the first class, we have two key laws~(given in \figname\ref{fig:SplitDistrTB}) for the decomposition and narrowing of time budgets. Whereas the first \lawname\ref{law:SplitBudget} replaces a single time budget by a sequence of two time budgets, the second \lawname\ref{law:NarrowBudget} incurs a reduction of nondeterminism that narrows a time budget. A point of design in applying these laws is to decide on the values of $t_1$ and $t_2$, which subsequently determine the amount of time available to the underlying data operations. Decomposition may, of course, be applied iteratively, so that a single time budget can be split into several time budgets for any given number of operations.

\begin{figure}[t]
\centering
\fbox{
\begin{minipage}{0.81\textwidth}
\begin{law}
$\;\circwait ~ 0 \upto t \; \equiv \; \circwait ~ 0 \upto t_1 \circseq \circwait ~ 0 \upto t_2 \quad \mbox{where} \quad t = t_1 + t_2$
\label{law:SplitBudget}
\end{law}
\vspace{-1.4em}
\begin{law}
$\;\circwait ~ 0 \upto t_1 \; \refby \; \circwait ~ 0 \upto t_2 \quad \mbox{where} \quad t_2 \leq t_1$
\label{law:NarrowBudget}
\end{law}
\vspace{-1.4em}
\begin{law}
Assuming $Op$ is a data operation and $P$ is a {\Circus} process, we have
$P (\circwait ~ t_1 \upto t_2 \circseq Op) \; \equiv \; P (Op \circseq \circwait ~ t_1 \upto t_2)$
\label{law:DistrBudget}
\end{law}
\end{minipage}}
\caption{Laws for decomposition and distribution of time budgets.}
\label{fig:SplitDistrTB}
\end{figure}

The second class of laws addresses the issue of moving the decomposed time budgets to suitable locations to attach them to their respective data operations. For this, we first transform all Z schema compositions into {\Circus} action sequences. The standard law for this is recaptured below from~\cite{ZRC}.
\begin{law}
$Op_1 \semi Op_2 \; \equiv Op_1 \circseq \! Op_2 \;\; \mbox{provided} \;\; \pre(Op_1 \semi Op_2) \land Op_1 \implies \pre'(Op_2)$
\end{law}
As usual, $\pre(Op)$ yields the precondition~(domain) of a Z operation $Op$, and we use $\pre'(Op)$ to indicate that the variables in the result are primed. We note that the semicolon `$\semi$' is used for composition of Z operations, as opposed to `$\circseq\!$' which is used for composition of {\Circus} actions.

We further require the specialised distribution \lawname\ref{law:DistrBudget} in \figname\ref{fig:SplitDistrTB}. This law is in fact non-compositional: it is a law about processes rather than actions. Hence, it only holds if the underlying action $\circwait t_1 \upto t_2 \circseq \! Op$ is embedded in a process $P$. The justification for the law comes from the structure and semantics of processes that prevents observation of the precise time at which an (internal) state change takes place. A proof is possible by induction over the structure of processes.

We note that no distribution laws exist to move time budgets across prefixes, since such transformations would not be correct as they alter the observable behaviour. Consider, for example, $c \then \circwait ~ t \circseq A$. Refining this action by $\circwait ~ t \circseq c \then A$ would be wrong since the refining action refuses communication on the channel $c$ for $t$ time units, whereas the refined action offers it immediately. Some general laws for {\Circus} refinement in~\cite{Oli05} are useful, too, namely to distribute time budgets into and out of internal and external choice. Lastly, we have a fusion law for nondeterministic choice of time budgets:
\begin{law}
$\circwait t_1 \upto t_2 \; \intchoice \; \circwait t_1' \upto t_2' \; \equiv \; \circwait \, \min(t_1, t_1') \upto \max(t_2, t_2')$
\end{law}
This law is useful as it enables the combination of two budgets.

The laws we present here are evidently complete for mission specifications in which each abstract data operation is already associated with an (abstract) time budget. Automation of the refinement can be envisaged by annotating each data operation with the intended time budget, and using tactics to mechanically perform the decomposition and distribution steps. An overall caveat for the transformation is that we cannot distribute time budgets into parallel data operations which are represented by Z schema conjunctions. This is because the conjunction operator only applies to schemas and not to actions, and the schema calculus does not support timing constructs such as $\circwait ~ t_1 \upto t_2$. (In our strategy, we, therefore, distribute the budgets of parallel operations after the {\Circus} parallel operators are introduced.)

The next section examines the refinement of sequential actions and schema conjunctions, as they emerge from the laws discussed so far, into parallel actions.


\subsection{Introduction of parallel handler actions}
\label{sec:Stage3}

In Section~\ref{sec:Stage1}, we have presented laws to parallelise data operations using schema conjunction, but considered no laws to parallelise actions. The laws we discuss next can be used to parallelise mission actions. Like in Section~\ref{sec:Stage1}, we divide the necessary laws into two classes:~laws that account for sequential designs and laws that cater for parallel designs. The shapes we target are precisely those produced by earlier decomposition of data operations, which makes this aspect of the verification more susceptible to automation. In the sequel, we discuss both classes of laws.

\paragraph{Laws for sequential handler designs}

Two central laws for parallelisation of handlers are given in \figname\ref{fig:seq-to-par-1} and \figname\ref{fig:seq-to-par-2}. The first one assumes that there exists no data dependency between the sequential handler actions $A_1$ and $A_2$, hence we have the proviso $\wrt(A_1) \, \cap \, \used(A_2) = \emptyset$, which states that the state components written by $A_1$ are disjoint from those read by $A_2$. A fresh typeless channel $c$ is introduced to control the order of execution of the parallel actions:~they both have to synchronise on it, so that the right parallel action $c \then A_2$ blocks until the left parallel action is ready to execute the prefix $c \then \Skip$. The channel $c$ models an SCJ event that is bound to the second handler and fired by the first handler.

\begin{figure}[t]
\centering
\fbox{
\begin{minipage}{0.72\textwidth}
\begin{law}
Let $A_1$ and $A_2$ be actions and $c$ a fresh typeless channel. Then,
\vspace{-4pt}
\input{seq-to-par-1.tex}
\label{law:seq-to-par-1}
\end{law}
\vspace{-2.2em}
\end{minipage}}
\caption{Parallelisation of independent sequential data operations.}
\label{fig:seq-to-par-1}
\end{figure}

\begin{figure}[t]
\vspace{1em}
\centering
\fbox{
\begin{minipage}{0.77\textwidth}
\begin{law}
Let $A_1$ and $A_2$ be actions and $c$ a fresh channel. Then,
\vspace{-4pt}
\input{seq-to-par-2.tex}
\label{law:seq-to-par-2}
\end{law}
\vspace{-2.2em}
\end{minipage}}
\caption{Parallelisation of dependent sequential data operations.}
\label{fig:seq-to-par-2}
\end{figure}

The second law~(\figname\ref{fig:seq-to-par-2}) assumes that there is a data dependency between the sequential handlers. In that case, the channel $c$ is parametrised by the type of the data that is passed between $A_1$ and $A_2$. Multiple data items can be passed by using product types, and, as mentioned earlier, class types are permissible, too. An interesting observation at this point is that the channel $c$ fulfils a dual purpose:~it controls both the order of execution of handlers and makes available shared data. Further refinement is hence required to untangle these concerns, namely by way of encapsulating the shared data independently of the control aspect. This is, however, beyond the scope of parallelisation of handlers and a separate and orthogonal design issue, so we do not discuss it further here. The report~\cite{ZCWWW12} examines it in detail though.

We emphasise that the parallelisations performed by \lawname\ref{law:seq-to-par-1} and \lawname\ref{law:seq-to-par-2} are to align the model with the SCJ paradigm and architecture. In other words, they do not parallelise the computations of the respective handlers, which are still performed in sequence here. This reflects that any sequentialism in an SCJ design needs to be explicitly enforced, while parallel execution~(of handlers) is the default.

For multiple applications of the two laws, we also require the application of several elementary {\Circus} laws between each application of \lawname\ref{law:seq-to-par-1} or \lawname\ref{law:seq-to-par-2}\commentout{; the majority of them is included in \figname\ref{fig:ExtractHideLaws} and \figname\ref{fig:DistrPrefixLaws}}. Their purpose is firstly to extract the newly introduced channel $c$ to the outer level of the mission action in which the targeted (refined) action is embedded\commentout{~(\figname\ref{fig:ExtractHideLaws})}, and secondly to distribute prefixes $c~[?~x] \then A_2$ introduced in the right-hand parallel action into $A_2$, namely if $A_2$ is itself an action sequence or parallelism\commentout{~(\figname\ref{fig:DistrPrefixLaws})}.

We conclude by observing that the first parallelisation \lawname\ref{law:seq-to-par-1} targets precisely the shape of models generated by earlier application of \lawname\ref{law:SeqPattn1}~(\figname\ref{fig:SeqPattn1}), and the second parallelisation \lawname\ref{law:seq-to-par-2} precisely the shape of models generated by earlier application of \lawname\ref{law:SeqPattn2}~(\figname\ref{fig:SeqPattn2}), subsequent to replacing Z compositions by action sequences, which is done collaterally as part of the distribution of time budgets.\commentout{Although some intermediate refinement steps are necessary, namely via the laws in \figname\ref{fig:ExtractHideLaws} and \figname\ref{fig:DistrPrefixLaws}, it turns out that those steps are systematic, making automation indeed possible here.}

\paragraph{Laws for parallel handler designs}

A key law for transforming parallel data operations modelled by conjunctions into parallel actions is presented in \figname\ref{fig:conj-to-par-1}. It applies to data operations $Op_1$ and $Op_2$ that write to disjoint sets of variables, which is what we usually expect from a parallelism at that level.

Although this law permits us to replace parallel data operations by parallel actions, this might not immediately yield a top-level parallelism of handlers as present in our refinement target in \figname\ref{fig:Target}. It shows, in general, that due to the fact that the conjunction might be embedded into action sequences~(see \lawname\ref{law:ParPattn2}), there is still a considerable number of refinement steps and specialised laws required to arrive at the desired shape. In particular, these refinements involved further decomposition of time budgets related to the particular parallel design adopted. Applying \lawname\ref{law:conj-to-par-1}, for instance, to the result of \lawname\ref{law:ParPattn2}~(\figname\ref{fig:ParPattn2}), we observe that there still remains a sequential composition with $MOp$. We can parallelise it using \lawname\ref{law:seq-to-par-2} in the previous section, but this does not completely eliminate it due to a prefix emerging in the left parallel action. In~\cite{ZCWWW12}, we precisely detail the basic refinement steps that are needed prior and subsequent to application of \lawname\ref{fig:conj-to-par-1}; they involve two specialised laws:~one for channel decomposition and one for distribution of an interleaving of basic communications into a preceding parallelism.\commentana{We do not discuss them and the underlying refinement here for reasons of space, but emphasise that parallelisation via \lawname\ref{fig:conj-to-par-1} alone is feasible, but requires manual work that can pose a significant challenge to automation.}

\begin{figure}[t]
\centering
\fbox{
\begin{minipage}{0.58\textwidth}
\begin{law}
\input{conj-to-par-1}
\label{law:conj-to-par-1}
\end{law}
\end{minipage}}
\caption{Low-level law for refining parallel data operations into actions.}
\label{fig:conj-to-par-1}
\end{figure}

\begin{figure}[t]
\centering
\fbox{
\begin{minipage}{0.79\textwidth}
\begin{law}
\input{conj-to-par-2}
\label{law:conj-to-par-2}
\end{law}
\end{minipage}}
\caption{High-level law for refining parallel data operations into actions.}
\label{fig:conj-to-par-2}
\end{figure}

We also consider high-level parallelisation laws. Namely, \lawname\ref{law:conj-to-par-2} in \figname\ref{fig:conj-to-par-2} directly targets shapes emerging from parallelising data operation via \lawname\ref{law:ParPattn2} and at the same time caters for further decomposition of time budgets. This shows in the time budgets $POp_{TB}$, $Rec_{TB}$ and $Merge_{TB}$ replacing the global time budget $Op_{TB}$. We hence have a proviso $POp_{TB} + n*Rec_{TB} + Merge_{TB} \leq Op_{TB}$ that considers the time allowance of the parallelised operations to compute the partial results, the time to record them, and the time needed to merge them. The concrete value of these budgets has to be determined by the developer as part of the verification process.

A design artifact of \lawname\ref{law:conj-to-par-2} is that it introduces a fresh typed channel $rec$ that is used to communicate the partial results to a parallel operation that receives and merges them into the final result. From this, a control fragment emerges that is later refined into shared data to hold the partial result(s); it contributes to the $HdlControl$ action in~\figname\ref{fig:Target} and its refinement gives rise to further design of how partial results are stored and processed; this relies on its own set of laws which are omitted here.

To conclude this aspect of the refinement, we observe that we can either tackle it by way of applying the more general \lawname\ref{law:conj-to-par-1}, or use specialised high-level laws like \lawname\ref{law:conj-to-par-2} that encapsulate particular designs. Since it is still an open issue how the general case can profit from further elementary laws and their automation, we recommend the use of high-level laws. Therefore, we assume that for every decomposition law into a parallel data operation, there exists at least one specialised action law that directly targets the emerging shape. So far, this appears to be the case, however, further experience needs to be gained to ascertain this. In~\cite{ZCWWW12}, we sketch a proof of \lawname\ref{law:conj-to-par-2} which uses a few novel and interesting elementary laws. Beyond this, future work may propose alternative parallelisation laws with more sophisticated merge operations that can, for instance, deal with partial results of heterogeneous type. We next look at an example that illustrates the refinement of a realistic SCJ program.


\section{Example}
\label{sec:Example}

As an example, we consider the refinement of an action that models the behaviour of the collision detector~({\CDx} benchmark) in \cite{KHPPTV09}. The {\CDx} SCJ program consists of a single mission that periodically carries out the following tasks:~reading a set of aircraft positions from a radar device, calculating their predicted motions, and identifying the number of aircraft at risk of colliding due to their distances decreasing below a certain threshold. Whereas~\cite{KHPPTV09} provides a sequential implementation using a single handler, we have developed a parallel program by breaking down the mission design into seven handlers:~(1) a cyclic input handler that reads the next radar frame; (2) a reducer handler that performs a voxel-hashing algorithm, which partitions the space; (3)~four parallel detector handlers that carry out the detection work; and (4)~an output handler that communicates the result.

Our starting point is the abstract operation $ComputeCycle$ in \figname\ref{fig:CDxCycle}. It is embedded into an action that defines the cyclic mission behaviour, as specified below.
\begin{circusaction}
  CDxMission ~ \circdef ~ \circmu X \circspot
  \\
  \t1
  \circblockopen
    \circblockopen
      \circblockopen
        (next\_frame~?~frame \then ComputeCycle) \circdeadlinesync ~ ~ INP\_DL \circseq
        \\
        \circwait 0 \upto (FRAME\_PERIOD - INP\_DL - OUT\_DL) \circseq
        \\
        (output\_collisions~!~collisions \then \Skip) \circdeadlinesync ~ ~ OUT\_DL
      \circblockclose
      \\
      \interleave ~ \circwait FRAME\_PERIOD
    \circblockclose
    \circseq X
  \circblockclose
\end{circusaction}
The channel $next\_frame$~(of a type $Frame$ encoding radar frames) is used to read the next frame of aircraft positions, and $output\_collisions$~(of type $\num$) to output the detected number of collisions. Collisions are computed by $ComputeCycle$ and stored in a state component $collisions$. The constant $FRAME\_PERIOD$ determines the length of a cycle, and $INP\_DL$ and $OUT\_DL$ are deadlines on external communications. We observe that $ComputeCycle$ is equipped with a time budget $FRAME\_PERIOD - INP\_DL - OUT\_DL$, obtained by subtracting from the cycle time the maximal amount of time that the communications are permitted to take. Besides, $RawFrame$, $StateTable$ and $Partition$ are {\OhCircus} classes.

\begin{figure}[t]
\centering
\begin{minipage}{0.98\textwidth}
\input{ComputeCycle.tex}
\end{minipage}
\caption{Z operation specifying the cyclic mission behaviour of the {\CDx}.}
\label{fig:CDxCycle}
\end{figure}

We start by decomposing $ComputeCycle$ into sequences and conjunctions of data operations. This is done by applying \lawname\ref{law:SeqPattn2} three times, followed by an application of \lawname\ref{law:ParPattn2}. This is not trivial, however, since the $ComputeCycle$ operation contains further existentially quantified variables that either correspond to abstract model variables~($posns$ and $motions$) here arising from earlier data refinement, or local variables like $voxel\_map$, capturing the result of the voxel-hashing algorithm.\commentana{(The model variables $posns$ and $motions$, recording aircraft positions and motions, are related to the corresponding program variables $currentFrame$ and $state$ via the abstraction functions $F$ and $G$ whose precise definition we omit as it is not relevant for our purpose of illustration.)} These quantifiers either have to be eliminated using the one-point rule, or localised to predicates corresponding to single handlers.

Another issue that needs to be addressed is that the data flow is not always explicit in abstract operations specifying missions. In our SCJ program, for example, data is transmitted between the reducer handler that carries out the voxel-hashing, and the detector handlers that perform the detection. That is, the reducer handler writes to the component $work$ which determines how the computational work is split\commentout{between the handlers}, and this variable is also read by the detector handlers. In the operation, the last existential conjunct
\[
  \exists collset : \finset ~ (Aircraft \cross Aircraft) | collset = CalcCollisionSet(posns', motions') @
  \\
  \t1 (\# collset = 0 \land collisions' = 0) \lor (\# collset > 0 \land collisions' \geq (\# collset) \div 2)
\]
models the detector handlers, and we notice that the new value of $collisions$ is determined by the function $CalcCollisions(posns', motions')$ in terms of the abstract model variables. To reformulate it in terms of $work$ requires some \textit{ad hoc} refinements that appear to be difficult to automate by a machine. We skip further details and merely present the result of the decomposition.
\begin{circusaction}
  CDxMission ~ \circdef ~ \circmu X \circspot
  \\
  \t1
  \circblockopen
    \circblockopen
      \circblockopen
        \circblockopen
          next\_frame~?~frame ~ \then
          \\
          \circblockopen
            RecordFrame \circseq
            \\
            ReduceAndPartitionWork \circseq
            \\
            DetectCollisions
          \circblockclose
        \circblockclose
        \circdeadlinesync ~ ~ INP\_DL \circseq
        \\
        \circwait 0 \upto (FRAME\_PERIOD - INP\_DL - OUT\_DL) \circseq
        \\
        (output\_collisions~!~collisions \then \Skip) \circdeadlinesync ~ ~ OUT\_DL
      \circblockclose
      \\
      \interleave \circwait FRAME\_PERIOD
    \circblockclose
    \circseq X
  \circblockclose
\end{circusaction}
where the decomposed Z operations $CalcPartCollisions$, $SetCollisionsFromParts$ and $DetectCollisions$ can be found in Appendix~\ref{sec:DecompExample}. We next decompose and distribute the time budget between the newly introduced sequential operations. For this, we introduce the handler-specific time budgets $RF_{TB}$, $RPW_{TB}$ and $DC_{TB}$. This yields the following refinement.
\begin{circusaction}
  CDxMission ~ \circdef ~ \circmu X \circspot
  \\
  \t1
  \circblockopen
    \circblockopen
      next\_frame~?~frame ~ \then
      \\
      \circblockopen
        \circwait 0 \upto RF_{TB} \circseq RecordFrame \circseq
        \\
        \circwait 0 \upto RPW_{TB} \circseq
        \\
        ReduceAndPartitionWork \circseq
        \\
        \circwait 0 \upto DC_{TB} \circseq DetectCollisions
      \circblockclose
      \circdeadlinesync ~ ~ INP\_DL
      \\
      \dots
    \circblockclose
    \circseq X
  \circblockclose
\end{circusaction}
The time budget $DC_{TB}$ is further decomposed during the parallelisation of actions. For the last part of the refinement parallelising the action above, we refer to the detailed description in our technical report~\cite{ZCWWW12}, which is available from \url{http://www.cs.york.ac.uk/circus/publications/techreports/}. It entails applying \lawname\ref{law:seq-to-par-2} and \lawname\ref{law:conj-to-par-2}, and after a finalising transformation that uses elementary laws and can be automated, too, we obtain an action that has the shape in~\figname\ref{fig:Target}.


\section{Conclusion}
\label{sec:Conclusion}

We have presented a collection of {\Circus} refinement laws that can be used to refine sequential specifications of SCJ mission behaviour into parallel designs that match the SCJ Level 1 programming model. Our refined models are a suitable starting point for further refinement of shared data and control mechanisms. We have also highlighted challenges for automation:~they are, primarily, in the decomposition of sequential and parallel data operations, and to provide a repository of parallelisation laws, both at the level of data operations and actions, that deal with a wide spectrum of recurring program designs. Due to the novelty of SCJ, there are still open issues related to the designs that ought to be supported, and hence we do not claim completeness at this stage. On the other hand, our results showed that the decomposition of time budgets can largely be automated, and so can (the intermediate steps in) the refinement of data operations into parallel handler actions, which ultimately creates a positive outlook. Like in SCJ, our model and strategy also supports data being shared between missions. But this is less of an issue for the refinement laws because no write conflicts or race conditions can arise. The mission design in fact emerges where sequential actions of an abstract centralised model are retained during refinement.

In practical terms, we propose to facilitate the decomposition of data operations, the more difficult aspect of a refinement, by asking the developer to identify intermediate target models that permit the application of one of the decomposition laws. Each intermediate model generates a refinement proof obligation which can be tackled in isolation, and, as we hope, its resolution will be able to take some advantage of automatic refinement tactics. The development of useful tactics is still work in progress, however, their mechanisation may use a tool like~\cite{ZOC12} to ensure soundness of refinements and laws alike.

An open issue is the validation of our laws against a semantics for the particular combination of {\Circus} languages that we use. Our recent work explores in detail the semantics of {\CircusTime}, and this shall provide a platform to prove, for instance, the laws about time budgets in Section~\ref{sec:Stage2}. Further work is, however, required to integrate that semantics with that of {\OhCircus}. And importantly, we require a proof that the laws from either language~({\OhCircus} and {\CircusTime}) hold within the combined language. The Unifying Theories of Programming~(UTP)~\cite{HH98}, the common semantic foundation for all {\Circus} dialects, ought to facilitate such a proof. It is an issue that is high on our agenda of research.

Related work includes action systems and their refinement~\cite{Back90,BK83}. Action systems combine state and behaviour by away of atomic \emph{actions} that operate on the state and that can be executed concurrently if there are no write conflicts to variables. Like {\Circus}, action systems come with an extensive refinement calculus, supporting the refinement of centralised sequential specifications into distributed implementations~\cite{Back90,BW03}. The computational paradigm is, however, more restrictive since actions have to adhere to a specific form, whereas {\Circus} actions can, for instance, use all of CSP's constructs.

Event-B~\cite{Abr11} is a practically-oriented formalism closely-related to action systems; it has been successfully used in the formal development of distributed systems in academia and industry. Research has been prompted to overcome initial restrictions of the method to deal with decomposition~\cite{But09} and time~\cite{CMR06}. It would be interesting to see whether Event-B would be expressive enough for SCJ handler models, and whether the refinement laws we propose can be formulated and perhaps validated.

SCJ is still a very recent technology, and, as far as we know, this is the first work that looks at refinement more specifically in the context of the SCJ programming model. Our results though contribute to a wider objective of proposing and proving refinement laws for \emph{all} aspects of the verification of SCJ programs. These are, among others, data refinements in {\CircusTime} and the introduction of class objects, the refinement of shared data and use of object references, and the transformation of models into {\SCJCircus}, a new language sufficiently concrete to be directly translatable into code. They are all immediate areas for future work, each bringing its own set of challenges for refinement and automation.
\vspace{-1em}


\paragraph{Acknowledgements}

This work was funded by the EPSRC grant EP/H017461/1. We are grateful to Andy Wellings for many useful clarifications of SCJ, and we also thank the anonymous reviewers for their pertinent and useful suggestions.


\bibliographystyle{eptcs}

\bibliography{refine2013_zeyda.bib}


\appendix

\section[Decomposed data operations of the CDx example]{Decomposed data operations of the {\CDx} example}
\label{sec:DecompExample}

\begin{schema}{CalcPartCollisions}
  \Xi ~ [currentFrame : RawFrame; state : StateTable; work : Partition; collisions : \num]
  \\
  i? : 1 \upto 4
  \\
  pcolls! : \num
\where
  pcolls! = \# ~ \{a_1 : Aircraft; a_2 : Aircraft | \exists l : work~.~getDetectorWork(i?).~elems~() @ \dots \} \div 2
\end{schema}
\begin{schema}{SetCollisionsFromParts}
  \Delta ~ [currentFrame : RawFrame; state : StateTable; work : Partition; collisions : \num]
  \\
  collsbag? : \bag ~ int
\where
  currentFrame' = currentFrame \land state' = state \land voxel\_map' = voxel\_map \land work' = work\\
  \exists s : \seq ~ int | s = items ~ collsbag? @ collisions' = \Sigma ~ s
\end{schema}
\begin{circusaction}
  DetectCollisions ~ \circdef ~\\
  \t1
  \circblockopen
    \circvar colls1, colls2, colls3, colls4 : \num \circspot\\
    \t1
    \circblockopen
      \lschexpract
        (\exists i? : \num @ CalcPartCollisions[colls1 / pcolls!] \land i? = 1) \land\\
        (\exists i? : \num @ CalcPartCollisions[colls2 / pcolls!] \land i? = 2) \land\\
        (\exists i? : \num @ CalcPartCollisions[colls3 / pcolls!] \land i? = 3) \land\\
        (\exists i? : \num @ CalcPartCollisions[colls4 / pcolls!] \land i? = 4)
      \rschexpract
    \circblockclose
    \circseq
    \\
    \t1 SetCollisionsFromParts(\lbag colls1, colls2, colls3, colls4 \rbag)
  \circblockclose
\end{circusaction}

\end{document}

%% file: Target.tex
\setlength{\zedindent}{0pt}
\begin{circusflow}
\begin{circus}
\circprocess ~ SCJDesign ~ \circdef ~ \circbegin
\end{circus}
\begin{circusaction}
  \circstate ~ State ~ == ~ [ ~ c_1 : T_1; c_2 : T_2, \dots, c_n : T_n |  Inv~(c_1, c_2, \dots, c_n) ~ ]
\end{circusaction}
\begin{circusaction}
  Init ~ == ~ [ ~ State' | InitOp~(c_1', c_2', \dots, c_n') ~ ]
\end{circusaction}
\begin{circusaction}
  Mission_i ~ \circdef ~
  \circblockopen
    \circblockopen
      Handler_1
      \\
      \lpar ns_1 | cs_1 \cup \lchanset termMsn \rchanset | ns_2 \cup \dots \cup ns_h \rpar
      \\
      Handler_2
      \\
      \lpar ns_2 | cs_2 \cup \lchanset termMsn \rchanset | ns_3 \cup \dots \cup ns_h \rpar
      \\
      \dots
      \\
      \lpar ns_{h-1} | cs_{h-1} \cup \lchanset termMsn \rchanset | ns_h \rpar
      \\
      Handler_h
    \circblockclose
    \\
    \lpar ns_1 \cup \dots \cup ns_h | cs_c \cup \lchanset termReq, termMsn \rchanset | ns_c \rpar HdlControl
  \circblockclose
\end{circusaction}
\begin{circusaction}
  \commentout{Mission_2 ~ \circdef ~ \dots \;\;} \mbox{where $Handler_i$ may be either defined as}
  \vspace{-5pt}
\end{circusaction}
\end{circusflow}
\begin{minipage}{0.52\textwidth}
\vspace{-15pt}
\begin{circusaction}
  Handler_i ~ \circdef ~ ~ \mbox{(for an aperiodic handler)}
  \\
  \t1
  \circblockopen \! \!
    \circmu X \circspot
    \circblockopen \!
      \circblockopen
        e_{(i, 1)}~?~v \then A(v) ~ ~ \extchoice
        \\
        e_{(i, 2)}~?~v \then A(v) ~ ~ \extchoice
        \\
        \dots
      \circblockclose
      \circseq \! X \!
      \\
      \extchoice ~ ~ termMsn \then \Skip
    \! \circblockclose
  \! \! \! \circblockclose
\end{circusaction}
\end{minipage}
$\,$ or $\,$
\begin{minipage}{0.42\textwidth}
\vspace{-41pt}
\begin{circusaction}
  Handler_i ~ \circdef ~ ~ \mbox{(for a periodic handler)}
  \\
  \t1
  \circblockopen \! \!
    \circmu X \circspot
    \circblockopen \!
      (A \circdeadlineterm T \interleave \circwait \, T) \circseq X
      \\
      \extchoice ~ ~ termMsn \then \Skip
    \! \circblockclose
  \! \! \! \circblockclose
\end{circusaction}
\end{minipage}
\vspace{-1pt}
\begin{circusflow}
\begin{circusaction}
  HdlControl ~ = ~ (termReq \then termMsn \then \Skip) \parallel ~ \dots
\end{circusaction}
\begin{circusaction}
  \circspot Init \circseq Mission_1 \circseq Mission_2 \circseq \dots \circseq Mission_m
\end{circusaction}
\begin{circus}
  \circend
\end{circus}
\end{circusflow}

%% file: SeqPattn1.tex
Let
$State ~ == ~ [ ~ x : T_1; y : T_2 | I_1(x) \land I_2(y) ~ ]$.
Then,\\[0.7em]
\begin{minipage}{0.31\textwidth}
\vspace{-15pt}
\setlength{\zedindent}{0pt}
\begin{schema}{Op}
  \Delta ~ State
\where
  P(x, y, x') \land Q(y, y')
\end{schema}
\end{minipage}
$\;\; \equiv \;\;$
\begin{minipage}{0.24\textwidth}
\vspace{-15pt}
\setlength{\zedindent}{0pt}
\begin{schema}{Op_1}
  \Delta ~ [ ~ x : T_1 | I_1(x) ~ ]
  \\
  \Xi ~ [ ~ y : T_2 | I_2(y) ~ ]
\where
  P(x, y, x')
\end{schema}
\end{minipage}
$\;\; \semi \;\;$
\begin{minipage}{0.24\textwidth}
\vspace{-15pt}
\setlength{\zedindent}{0pt}
\begin{schema}{Op_2}
  \Delta ~ [ ~ y : T_2 | I_2(y) ~ ]
  \\
  \Xi ~ [ ~ x : T_1 | I_1(x) ~ ]
\where
  Q(y, y')
\end{schema}
\end{minipage}

%% file: SeqPattn2.tex
Let
$State ~ == ~ [ ~ x : T_1; y : T_2 | I_1(x) \land I_2(x, y) ~ ]$.
Then,\\[0.7em]
\begin{minipage}{0.34\textwidth}
\vspace{-15pt}
\setlength{\zedindent}{0pt}
\begin{schema}{Op}
  \Delta ~ State
\where
  P(x, y, x') \land Q(x', y, y')
\end{schema}
\end{minipage}
$\;\; \equiv \;\;$
\begin{minipage}{0.25\textwidth}
\vspace{-15pt}
\setlength{\zedindent}{0pt}
\begin{schema}{Op_1}
  \Delta ~ [ ~ x : T_1 | I_1(x) ~ ]
  \\
  \Xi ~ [ ~ y : T_2 ~ ]
\where
  I_2(x, y) \land
  \\
  P(x, y, x')
\end{schema}
\end{minipage}
$\;\; \semi \;\;$
\begin{minipage}{0.25\textwidth}
\vspace{-15pt}
\setlength{\zedindent}{0pt}
\begin{schema}{Op_2}
  \Xi ~ [ ~ x : T_1 | I_1(x)~ ]
  \\
  \Delta ~ [ ~ y : T_2 ~ ]
\where
  I_2(x', y') \land
  \\
  Q(x, y, y')
\end{schema}
\end{minipage}

%% file: ParPattn2.tex
Let
$State ~ == ~ [ ~ x : T_1; r : T_2 | I_1(x) \land I_2(x, r) ~ ]$.
Then,\\[0.7em]
\begin{minipage}{0.32\textwidth}
\vspace{-15pt}
\setlength{\zedindent}{0pt}
\begin{schema}{Op}
  \Xi ~ [ ~ x : T_1 | I_1(x) ~ ]
  \\
  \Delta ~ [ ~ r : T_2 | I_2(x, r) ~ ]
\where
  \exists r_1, \dots, r_n : T_2 |
  \\
  \circblockopen
    Q(r_1, 1, x\commentout{, r}) \land
    \\
    Q(r_2, 2, x\commentout{, r})  \land
    \\
    \dots
    \\
    Q(r_n, n, x\commentout{, r})
  \circblockclose
  @
  \\
  r' = r_1 ~ \mathbin{op} ~ r_2 ~ \mathbin{op} ~ \dots ~ \mathbin{op} ~ r_n
\end{schema}
\end{minipage}
\vspace{1em}
$\quad \equiv \quad$
\begin{minipage}{0.46\textwidth}
\vspace{-15pt}
\setlength{\zedindent}{0pt}
\begin{circusaction}
  \circblockopen
    \circvar r_1, \dots, r_n : T_2 \circspot
    \\
    (\exists i? : \num @ POp[r_1 / r!] \land i? = 1) \land
    \\
    (\exists i? : \num @ POp[r_2 / r!] \land i? = 2) \land
    \\
    \dots
    \\
    (\exists i? : \num @ POp[r_n / r!] \land i? = n) \circseq
    \\
    MOp(\lbag r_1, \dots, r_n \rbag)
  \circblockclose
\end{circusaction}
\end{minipage}
\par
where $\;$
\begin{minipage}{0.24\textwidth}
\vspace{-15pt}
\setlength{\zedindent}{0pt}
\begin{schema}{POp}
  \Xi ~ [ ~ x : T_1 | I_1(x) ~ ]
  \\
  r! : T_2
  \\
  i? : 1 \upto n
\where
  Q(r!, i?, x\commentout{, r})
\end{schema}
\end{minipage}
$\;\;$ and $\;\;$
\begin{minipage}{0.36\textwidth}
\vspace{-15pt}
\setlength{\zedindent}{0pt}
\begin{schema}{MOp}
  \Xi ~ [ ~ x : T_1 | I_1(x) ~ ]
  \\
  \Delta ~ [ ~ r : T_2 | I_2(x, r) ~ ]
  \\
  rb? : \bag ~ T_2
\where
  \exists s : \seq ~ T_2 | s = \mathop{items} rb? @
  \\
  r' = \mathbf{fold} \; op \; zero \; s
\end{schema}
\end{minipage}\\[7pt]
provided that $op$ is an associative and commutative binary operation. The function $\mathbf{fold}$ is the standard folding operation over a sequence of values and $zero$ a zero for $op$.

%% file: seq-to-par-1.tex
\setlength{\zedindent}{0pt}
\setlength{\zedskip}{2pt}
\begin{circusflow}
\begin{circusaction}
  A_1 \circseq A_2 ~ \equiv ~ ((A_1 \circseq c \then \Skip) \lpar \wrt(A_1) | \lchanset c \rchanset | \wrt(A_2) \rpar (c \then A_2)) ~ \circhide ~ \lchanset c \rchanset
  \also
  \provided \; \wrt(A_1) \cap \wrt(A_2) = \emptyset \;\; \mbox{and} \;\; \wrt(A_1) \cap \used(A_2) = \emptyset
\end{circusaction}
\end{circusflow}

%% file: seq-to-par-2.tex
\setlength{\zedindent}{0pt}
\setlength{\zedskip}{2pt}
\begin{circusflow}
\begin{circusaction}
  A_1 \circseq A_2 ~ \equiv ~ ((A_1 \circseq c~!~x \then \Skip) \lpar \wrt(A_1) | \lchanset c \rchanset | \wrt(A_2) \rpar (c~?~x \then A_2)) ~ \circhide ~ \lchanset c \rchanset
  \also
  \provided \; \wrt(A_1) \cap \wrt(A_2) = \emptyset \;\; \mbox{and} \;\; \wrt(A_1) \cap \used(A_2) = \{x\}
\end{circusaction}
\end{circusflow}

%% file: conj-to-par-1.tex
\begin{math}
  Op_1 \land Op_2 \; \equiv \; Op_1 \lpar \wrt(Op_1) | \emptyset | \wrt(Op_2) \rpar Op_2
  \\
  \provided \; \wrt(Op_1) ~ \cap ~ \wrt(Op_2) = \emptyset
\end{math}

%% file: conj-to-par-2.tex
\begin{math}
  \circwait ~ 0 \upto Op_{TB}
  \circseq
  \fbox{\mbox{RHS of \lawname\ref{law:ParPattn2}}}
  \;\; \refby
\end{math}
\\
\begin{math}
\circblockopen
  \circblockopen
    (\circvar r_1 : T \circspot \circwait ~ 0 \upto POp_{TB} \circseq (\exists i? : \num @ i? = 1) \circseq rec~!~r_1 \then \Skip) ~ \parallel\\
    (\circvar r_2 : T \circspot \circwait ~ 0 \upto POp_{TB} \circseq (\exists i? : \num @ i? = 2) \circseq rec~!~r_2 \then \Skip) ~ \parallel\\
    \dots\\
    (\circvar r_n : T \circspot \circwait ~ 0 \upto POp_{TB} \circseq (\exists i? : \num @ i? = n) \circseq rec~!~r_n \then \Skip)
  \circblockclose
  \\
  \t1 \lpar \emptyset | \lchanset rec \rchanset | \{r\} \rpar
  \\
  \circblockopen
    \circvar r_1, r_2, \dots, r_n : T \circspot
    \\
    \circblockopen
      (rec~?~x \then \circwait ~ 0 \upto Rec_{TB} \circseq r_1 := x)
      \\
      (rec~?~x \then \circwait ~ 0 \upto Rec_{TB} \circseq r_2 := x)
      \\
      \dots
      \\
      (rec~?~x \then \circwait ~ 0 \upto Rec_{TB} \circseq r_n := x)
    \circblockclose
    \circseq
    \\
    \circwait ~ 0 \upto Merge_{TB} \circseq MOp(\lbag r_1, r_2, \dots, r_n \rbag)
  \circblockclose
\circblockclose
\end{math}
$\provided POp_{TB} + n*Rec_{TB} + Merge_{TB} \leq Op_{TB}$

%% file: ComputeCycle.tex
\setlength{\zedindent}{0pt}
\begin{minipage}{0.97\textwidth}
\vspace{-1em}
\begin{schema}{ComputeCycle}
  \Delta ~ [currentFrame : RawFrame; state : StateTable; work : Partition; collisions : \num]\\
  frame? : Frame
\where
  \exists posns, posns', motions, motions' : Frame |
  \\
  \t1 \dom ~ posns = \dom ~ motions \land \dom ~ posns' = \dom ~ motions' @
  \\
  \exists voxel\_map : HashMap[Vector2d,List[Motion]] | voxel\_map \neq \circnull @
  \\
  \circblockopen
    posns' = frame? \land
    \\
    motions' = (\lambda a : \dom ~ posns' @ \IF a \in \dom ~ posns \THEN (posns' ~ a) \vminus (posns ~ a) \ELSE ZeroV) \land
    \\
    posns = F(currentFrame) \land motions = G(currentFrame, state) \land
    \\
    posns' = F(currentFrame') \land motions' = G(currentFrame', state') \land
    \\
    \circblockopen
      \forall a_1, a_2 : Aircraft | \{a_1, a_2\} \subseteq \dom ~ posns' @
      \\
      (a_1, a_2) \in CalcCollisionSet(posns', motions') \implies
      \\
      \t1
      \circblockopen \! \!
        \exists l : List[Motion] | l \in voxel\_map~.~values()~.~elems() @
        \\
        \mbox{``predicate that states the collision pair $(a_1, a_2)$ is in $l$''}
      \! \circblockclose
    \circblockclose
    \land
    \\
    voxel\_map~.~values()~.~elems() = \bigcup ~ \{i : 1 \upto 4 @ work'~.~getDetectorWork(i)~.~elems()\} \land
    \\
    \exists collset : \finset ~ (Aircraft \cross Aircraft) | collset = CalcCollisionSet(posns', motions') @
    \\
    \t1 (\# collset = 0 \land collisions' = 0) \lor (\# collset > 0 \land collisions' \geq (\# collset) \div 2)
  \circblockclose
\end{schema}
\end{minipage}